\documentclass[preprint2]{aastex62}
\submitjournal{ApJ}
\usepackage{natbib}

\newcommand{\sups}[1]{\textsuperscript{#1}}
\newcommand{\subs}[1]{$_{#1}$}

\shorttitle{Follow-up Timing of Three GMRT Pulsars}
\shortauthors{Surnis et al.}

\begin{document} 

\title{Follow-up Timing of Three GMRT Pulsars}

\correspondingauthor{Mayuresh P. Surnis}
\email{mpsurnis@ncra.tifr.res.in, mayuresh.surnis@mail.wvu.edu}

\author[0000-0002-9507-6985]{Mayuresh. P. Surnis}
\affil{National Centre for Radio Astrophysics, Tata Institute of Fundamental Research, P. O. Bag 3, University of Pune Campus, Ganeshkhind, Pune, India.}
\affiliation{Radio Astronomy Centre, National Centre for Radio Astrophysics, Tata Institute of Fundamental Research, Udhagamandalam, India.}
\affiliation{West Virginia University, Department of Physics and Astronomy, P. O. Box 6315, Morgantown, WV, USA.}
\affiliation{Center for Gravitational Waves and Cosmology, West Virginia University, Chestnut Ridge Research Building, Morgantown, WV, USA}

\author[0000-0002-0863-7781]{Bhal Chandra Joshi}
\affil{National Centre for Radio Astrophysics, Tata Institute of Fundamental Research, P. O. Bag 3, University of Pune Campus, Ganeshkhind, Pune, India.}

\author[0000-0001-7697-7422]{Maura A. McLaughlin}
\affil{West Virginia University, Department of Physics and Astronomy, P. O. Box 6315, Morgantown, WV, USA.}
\affiliation{Center for Gravitational Waves and Cosmology, West Virginia University, Chestnut Ridge Research Building, Morgantown, WV, USA}

\author[0000-0003-4528-2745]{M. A. Krishnakumar}
\affil{Radio Astronomy Centre, National Centre for Radio Astrophysics, Tata Institute of Fundamental Research, Udhagamandalam, India.}

\author[0000-0003-4274-211X]{P. K. Manoharan}
\affil{Radio Astronomy Centre, National Centre for Radio Astrophysics, Tata Institute of Fundamental Research, Udhagamandalam, India.}

\author[0000-0002-9225-9428]{Arun Naidu}
\affil{National Centre for Radio Astrophysics, Tata Institute of Fundamental Research, P. O. Bag 3, University of Pune Campus, Ganeshkhind, Pune, India.}

\begin{abstract}
We report on the results of multi-frequency follow-up observations of three pulsars (PSRs J0026+6320, J2208+5500 and J2217+5733) discovered with the Giant Metrewave Radio Telescope (GMRT). These observations were carried out with the GMRT and the Ooty Radio Telescope (ORT). We report improved timing solutions for all three pulsars. For PSR J2208+5500, we estimate the nulling fraction to be 53(3)\%. The steep spectrum of this pulsar, its single component profile, and narrow pulse width suggest its single component to be a core component. If so, this significant cessation of emission in a core component is inconsistent with a geometric origin of nulls, such as those due to `empty' sightline traverses, and more likely due to intrinsic changes in the pulsar magnetosphere. We have measured scatter-broadening timescales at 325 and 610 MHz for PSRs J0026+6320 and J2217+5733. The implied scatter-broadening frequency scaling index of $-$2.9 for both pulsars is different from that expected assuming Kolmogorov turbulence in the interstellar medium. We also report spectral indices, obtained from imaging observations, for all three pulsars for the first time. The spectra for two of these pulsars indicate a possible spectral turnover between 100$-$300 MHz. Multi-frequency timing analyses carried out for these pulsars have enabled us to determine dispersion measures (DMs) with accuracies of 0.01 pc cm\sups{--3}. This demonstrates the usefulness of quasi-simultaneous multi-frequency multi-epoch timing observations with the GMRT and the ORT for studying variations in DM for millisecond pulsars.

\end{abstract}

\keywords{(stars:) pulsars: general -- (stars:) pulsars: individual (PSR J0026+6320, J2208+5500, PSR J2217+5733)}

\section{Introduction}
\label{intro}
Dedicated timing observations help in constraining many pulsar parameters like the period (P), period derivative (\.{P}), and the position of the pulsar. Multi-frequency timing observations further help in an accurate determination of the dispersion measure (DM). High cadence multi-frequency timing observations can be used to track changes in DM, which is important for high precision pulsar timing, such as the pulsar timing array experiment \citep{yhc+07a,yhc+07b,mhb+13,kcs+13,lcc+16,bkk+16,jml+17}. With an interferometric array like the Giant Metrewave Radio Telescope (GMRT), simultaneous imaging data at different frequencies is also useful to constrain the spectrum of a pulsar, particularly for the pulsars showing highly scatter-broadened pulse profiles [See \cite{dbk+15}]. Additionally, imaging data help in understanding the environment of the pulsar. Regular multi-frequency monitoring observations are also useful in monitoring long term flux density variations and in the study of pulsar emission phenomena like nulling, mode changing, scatter-broadening and profile evolution.

In a blind survey carried out with the GMRT at 610 MHz, \cite{jml+09} discovered three pulsars, PSRs J0026+6320, J2208+5500, and J2217+5733. The initial follow-up timing observations of these pulsars were carried out with the GMRT at 610 MHz and the Lovell telescope (Jodrell Bank) at 1400 MHz. These pulsars belong to the normal pulsar population with rotation periods of 318, 933, and 1056 ms, respectively. \cite{jml+09} also reported on the nulling behavior of one of the pulsars, PSR J2208+5500, but could only put a lower limit of 7.5\% on its nulling fraction (NF) due to the limited signal to noise ratio (S/N) of single pulses in both 610 and 1400 MHz observations. Compared to the mean pulsar spectral index of $-$1.4, with variance of 1 \citep{blv13}, two of the three pulsars, PSRs J2208+5500 and J2217+5733, were found to have steep spectral indices of $-$2.0(0.9) and $-$2.7(1.3), respectively, while PSR J0026+6320 was found to have a much flatter spectral index of $-$0.8(0.3)\citep{jml+09}\footnote{All errors in parentheses are 1-sigma uncertainties, unless mentioned otherwise.}. PSRs J2217+5733 and J0026+6320 have high measured DMs and therefore are likely to show significant scatter-broadening at lower frequencies.

Since the discovery of these pulsars, we have continued the multi-frequency follow-up timing and radio imaging observations. In this paper, we present the results of these observations carried out with the GMRT and the Ooty Radio Telescope (ORT). Details of the observations are presented in Section \ref{obs}. The data analysis procedure is briefly explained in Section \ref{anal}. The results for each pulsar are presented in Section \ref{anal_res}. The implications of our results are discussed in Section \ref{disc}. 

\section{Observations}
\label{obs}
We carried out the observations with the GMRT (located 80 km North of Pune, India), which has 30 single dishes, each with diameter of 45 m, and an effective collecting area of 30,000 m\sups{2} and a gain of 0.32 K Jy\sups{--1} antenna\sups{--1} \citep{sak+91} and the ORT (located 10 km South-West of Ooty, India), which has a single polarization reflector with an effective collecting area of 8,700 m\sups{2} and a gain of 3.3 K Jy\sups{--1} \citep{njm+15}. The typical system temperature of the GMRT is about 100 K at 325 and 610 MHz, while that of the ORT is about 150 K at 326.5 MHz.

The timing observations were carried out with the GMRT at 325 MHz between MJD 56051$-$56745 with a cadence of about two weeks. Once a more accurate position was established, timing observations were subsequently moved to 610 MHz and were carried out between MJD 56955$-$57282 with a similar cadence. There were two epochs, MJDs 56983 and 57046, when we observed these pulsars at 1170 MHz to get better frequency coverage for determining DMs and spectral indices. The GMRT observations were carried out using the GMRT software backend \citep[GSB;][]{rgp+10} covering a bandwidth of 33 MHz, split into 512 spectral channels. The time-series data were recorded in the phased array (PA) mode, combining 16 closest antennae, with a sampling time of 123 $\mu$s, while the imaging data were simultaneously recorded with a sampling time of 16 s. The observations at the ORT for PSR J2208+5500 were started on MJD 56653 with almost daily cadence up to MJD 57234. Since then, the observations continued with a cadence of about twice a week up to MJD 57638. PSRs J0026+6320 and J2217+5733 were not detected at the ORT\footnote{This is due to the reduced sensitivity of the ORT at these declinations. Calibrator scans showed that the sensitivity is 35\% and 20\% of maximum, respectively, for PSR J0026+6320 and J2217+5733. This translates to an 8$\sigma$ sensitivity of 12 and 7 mJy for a 15 minute observation, respectively.}. The ORT data were recorded using the PONDER backend \citep{njm+15} in pulsar mode over a bandwidth of 16 MHz centered at 326.5 MHz with a sampling time of 0.5 ms. A known pulsar, PSR B1937+21, which has a well constrained timing solution, was also observed at each epoch with both telescopes as a test source.

\section{Data Analysis and Results}
\label{anal}

\subsection{Analysis of time-series data}
\label{anal_tim}
We began the GMRT data analysis with flagging the frequency channels suffering from radio frequency interference (RFI). We identified the RFI-affected channels by calculating the root-mean-square (rms) over the full time range for each frequency channel after subtracting a running mean every second. We masked the channels whose rms was more than three times the average rms across the band in the subsequent analysis. The resultant data were then analyzed using the SIGPROC\footnote{http://sigproc.sourceforge.net, http://adsabs.harvard.edu/abs/2011asc1.soft0701016L} \citep{sigproc} analysis package in the following manner. The data were first de-dispersed to the DM [reported by \cite{jml+09}] of the pulsar to produce time-series files, which were subsequently folded to get one time stamped integrated profile per epoch for each pulsar. At the ORT, the pulsar backend, with the help of polyco files produced by the pulsar timing analysis package TEMPO2\footnote{http://www.atnf.csiro.au/research/pulsar/tempo2/} \citep{hem06} produced coherently de-dispersed, time stamped integrated profiles [refer to \cite{njm+15} for details]. During the analysis of the GMRT data taken initially at 325 MHz, we noticed that PSR J2217+5733 was detected with a much lower S/N than the expected value and PSR J0026+6320 was detected at only some epochs. Hence, a local DM search was undertaken for both the pulsars. The range of DMs for the local search was 100$-$200 pc cm\sups{--3} for PSR J2217+5733 and 200$-$280 pc cm\sups{--3} for PSR J0026+6320. The local search resulted in improved DM estimates for both pulsars (243 pc cm\sups{--3} instead of 231 pc cm\sups{--3} for PSR J0026+6320 and 131 pc cm\sups{--3} instead of 162 pc cm\sups{--3} for PSR J2217+5733). All the GMRT data for these pulsars were subsequently de-dispersed to the corrected DMs for further analysis.  

The integrated profiles for each pulsar were added together, after aligning the peaks manually, to produce a high S/N ($>$100) average profile at each frequency. This was smoothed to generate a noise-free template profile, which was used to estimate the pulse time of arrival (TOA) at each epoch using the template matching technique as described by \cite{t92}. The timing analysis was done using the pulsar timing package TEMPO2 \citep{hem06}. The average profiles obtained for all the pulsars are shown in Figures 1 -- 3 (note that the flux density calibration for all the average profiles was carried out using the radiometer equation). We used a multiplicative factor (EFAC) to scale the TOA uncertainties so as to get realistic error bars on the parameters of the timing solution. The value of EFAC was estimated independently for TOAs at each telescope and frequency so as to obtain a reduced $\chi$\sups{2} of 1 (see  Table \ref{anal_tim_efac}). At the ORT, the initial configuration of the time and frequency system introduced a 300 $\mu$s jitter in TOA measurements. Once this problem was identified, changing the clock configuration reduced the jitter drastically. Unfortunately there is no way to correct the data for this effect afterwards. We have still not excluded the ORT TOAs for PSR J2208+5500 from the timing analysis because the rms clock jitter is much smaller than the rms post-fit residual of PSR J2208+5500 ($\sim$ 1.2 ms).

\begin{table*}
\begin{center}
\begin{tabular}{cccccc}
\hline \hline
Telescope & Frequency & \multicolumn{3}{c}{Pulsar} \\
          & (MHz)     & J2208+5500 & J2217+5733 & J0026+6320 \\
\hline
ORT       & 325       & 1.2        & -          & -          \\
GMRT      & 325       & 4.2        & 6.5        & 3.5        \\
GMRT      & 610       & 5.2        & 4.8        & 3.8        \\
GMRT      & 1170      & 1.0        & 1.5        & -          \\
\hline
\end{tabular}
\caption{EFAC values for different telescopes and observing frequencies for all the observed pulsars. Dashes indicate unavailability of TOAs.}
\label{anal_tim_efac}
\end{center}
\end{table*}

As the observations were done at different telescopes and frequencies, constant phase offsets were determined in order to better estimate DMs. We did this by estimating the absolute instrumental offsets both at the ORT and the GMRT by radiating a noise signal modulated by the 1 pulse per minute (PPM) signal obtained from the GPS unit. The absolute instrumental delay (at both telescopes) was then estimated by subtracting the expected time of arrival of the rising edge of the signal from the observed time of arrival. The values were found to be 1258(0.5) ms for GSB and 490(0.5) ms for PONDER. The constant offset between GMRT$-$ORT was thus estimated to be 768(1) ms, while no offsets were seen at GMRT between 325$-$610 and 325$-$1170 MHz down to a level of 0.05 ms. The GMRT$-$ORT offset was found to be consistent for the TOAs of PSR J2208+5500. Given that the error in estimated instrumental offset, dominated by the jitter, is 1 ms, while the rms residuals obtained in all the pulsars are greater than 1 ms, the DM uncertainties obtained from the timing fit are more dominant than the ones obtained from the uncertainties in the offset. 

\begin{figure}
\begin{center}

\includegraphics[scale=0.7]{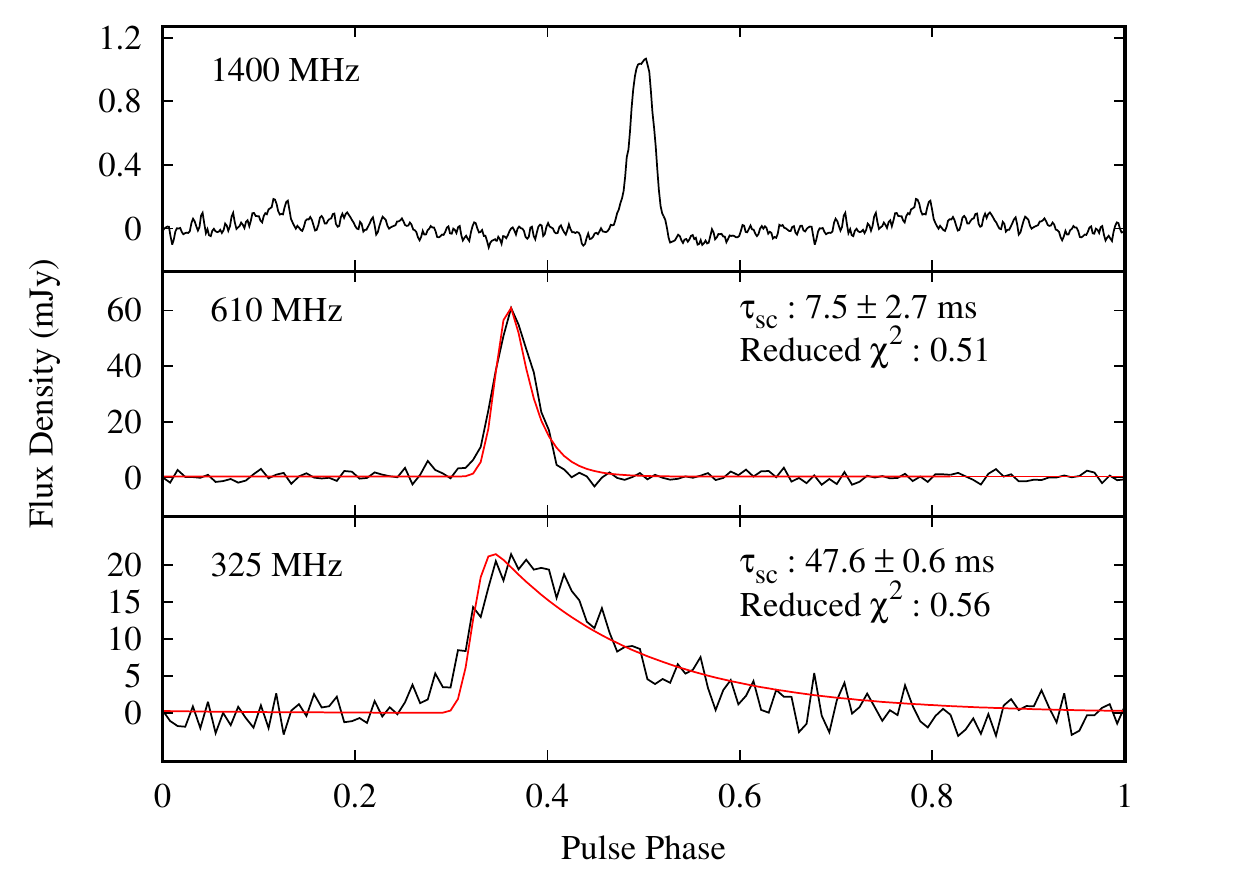}
\caption{Average profiles for PSR J0026+6320 at three observing frequencies. The top panel shows the average profile at 1400 MHz observed with the Lovell telescope (taken from \cite{jml+09}). The bottom two panels show the average profiles at 610 and 325 MHz with the GMRT. Observed profiles are shown in black and the best fit scatter-broadening model is overplotted  in red. $\tau$\subs{sc} estimates with the reduced $\chi$\sups{2} from the best fit model for 610 and 325 MHz are shown in the top right corner of each panel.}
\label{res_0026_scatt}

\end{center}
\end{figure}

\begin{figure}
\begin{center}

\includegraphics[scale=0.7]{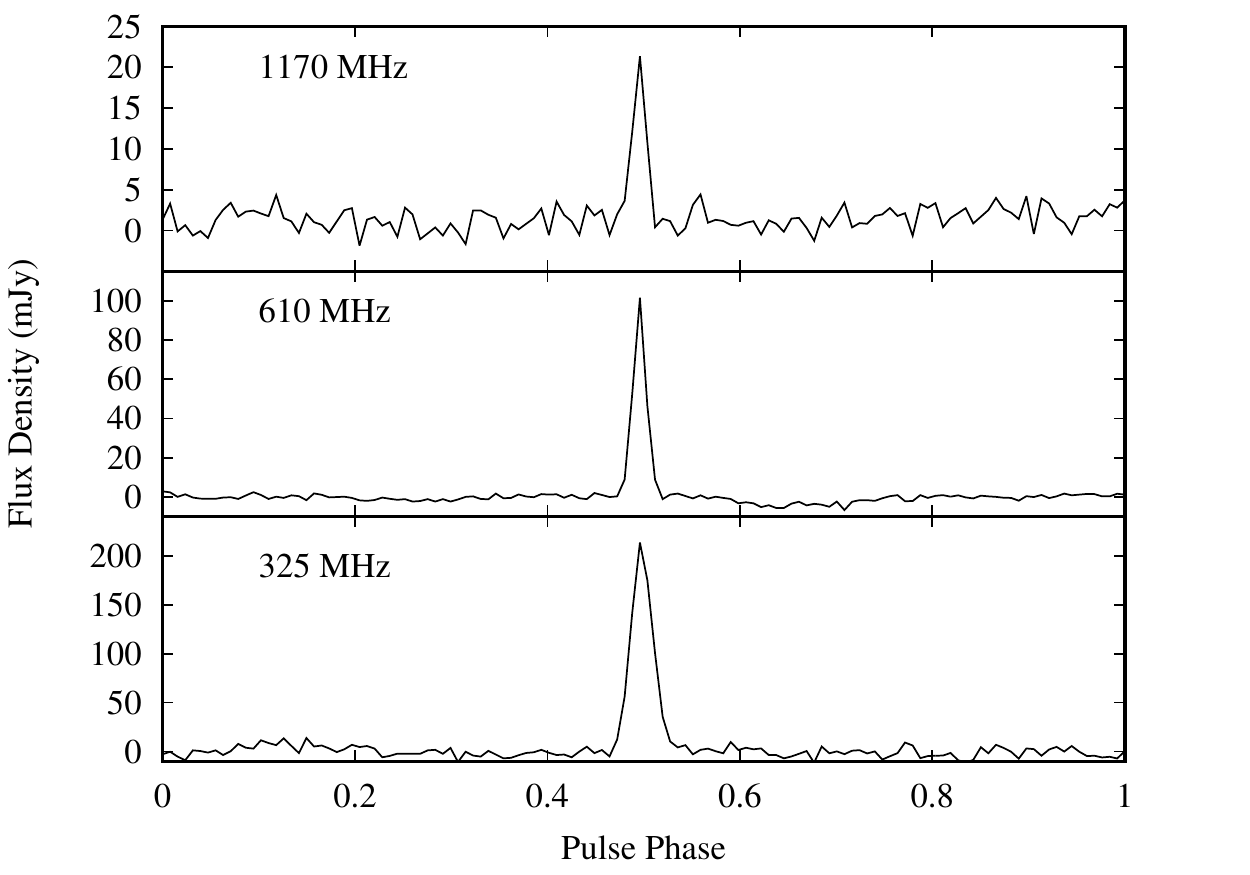}
\caption{Average profiles for PSR J2208+5500 at three observing frequencies with the GMRT.}
\label{anal_tim_2208}

\end{center}
\end{figure}

\begin{figure}
\begin{center}

\includegraphics[scale=0.7]{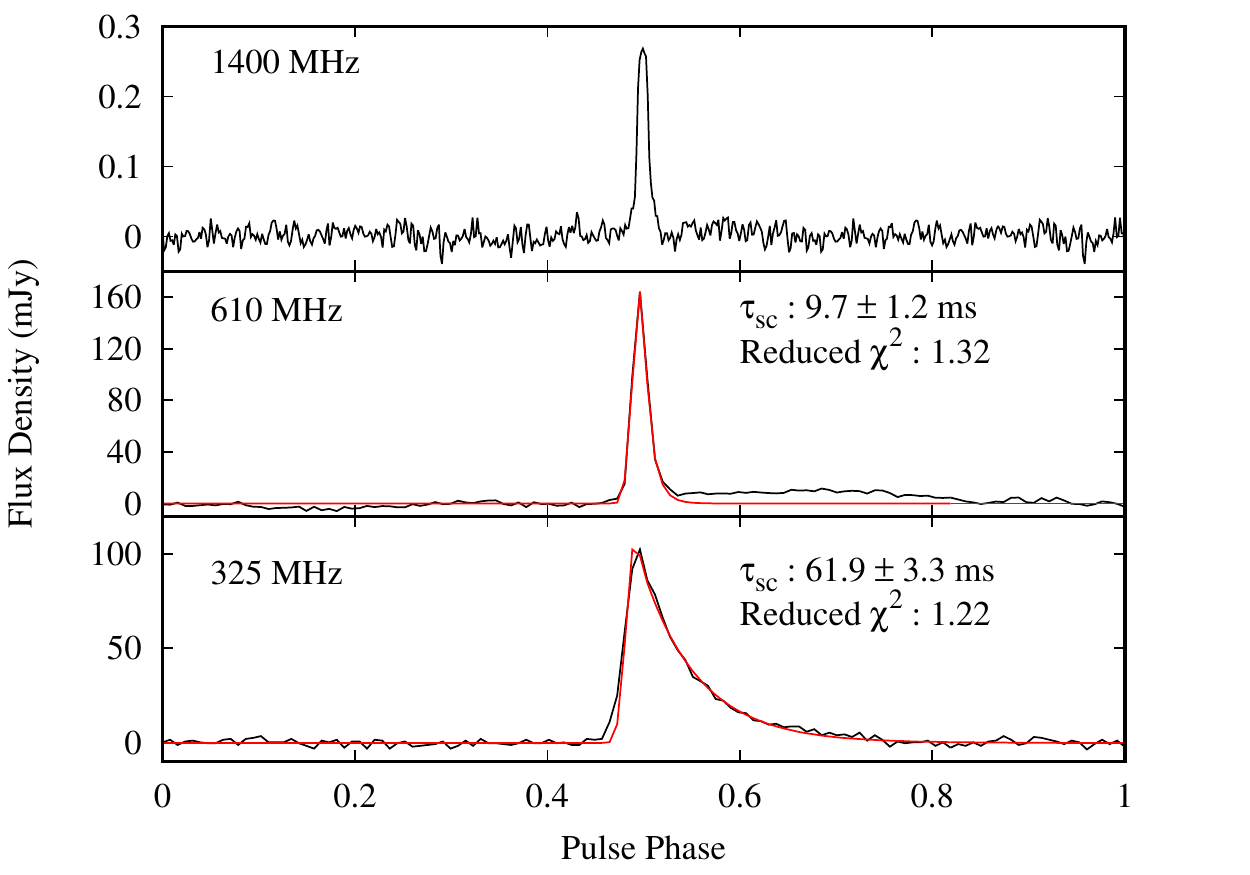}
\caption{Average profiles for PSR J2217+5733 at three observing frequencies. The top panel shows the average profile at 1400 MHz observed with the Lovell telescope (taken from \cite{jml+09}). The bottom two panels show the average profiles at 610 and 325 MHz with the GMRT. Observed profiles are shown in black and the best fit scatter-broadening model is overplotted  in red.  $\tau$\subs{sc} estimates with the reduced $\chi$\sups{2} from the best fit model for 610 and 325 MHz are shown in the top right corner of each panel.}
\label{res_2217_scatt}

\end{center}
\end{figure}

\subsection{Analysis of imaging data}
\label{anal_imag}

The imaging data, which were acquired simultaneously with time-series observations carried out at the GMRT, were flagged and calibrated using FLAGCAL\footnote{http://ncralib1.ncra.tifr.res.in:8080/jspui/handle/ 2301/581} \citep[e.g.,][]{c13}. At each epoch, 3C48 and 1822$-$096 were observed along with the pulsars. The former was used as flux calibrator to calibrate the raw visibilities and the latter was used for phase calibration. The images at 325 and 610 MHz were then made using the multi-facet imaging technique in the Astronomical Image Processing System (AIPS)\footnote{http://www.aips.nrao.edu/index.shtml} with 19 facets spread across the primary beam area. The images at 1170 MHz were made with the regular 2-D imaging technique. The image thus obtained was improved further with three rounds of self calibration with time scales for calculating the phase solutions of 4, 1, and 0.5 minutes. We noticed that the interferometric positions of the pulsars were slightly offset from the timing positions. The offsets were introduced due to the phase calibrator being more than 40\sups{$\circ$} away from the pulsars. We then compared the positions of other point sources in the primary beam to those given in the NRAO VLA Sky Survey \citep{ccg+98} catalog and confirmed that the offsets were systematic. The offsets were then corrected before further analysis. The flux densities were then measured by using the task JMFIT (from the calibrated images). This was done by fitting a single elliptical Gaussian function with width equal to the width of the synthesized beam for each epoch. The integrated flux within the best-fit Gaussian beam was taken as the flux density of the source. 
 
A search in the TIFR GMRT Sky Survey alternate data release \citep[TGSS ADR;][]{ijm+16}\footnote{http://tgssadr.strw.leidenuniv.nl/doku.php} resulted in no counterparts being detected for the pulsars under study. Based on the median rms of 3.5 mJy from \cite{ijm+16}, we estimate a 6$\sigma$ upper limit of 21 mJy on the flux densities of these pulsars. It should be noted that all TGSS flux density estimates are obtained from snap-shot images obtained in TGSS. The flux densities obtained using the radiometer equation were consistent with those obtained from imaging observations.

\begin{figure*}
\begin{center}

\includegraphics[trim=0.5cm 1.2cm 0 1.2cm, clip, scale=0.315]{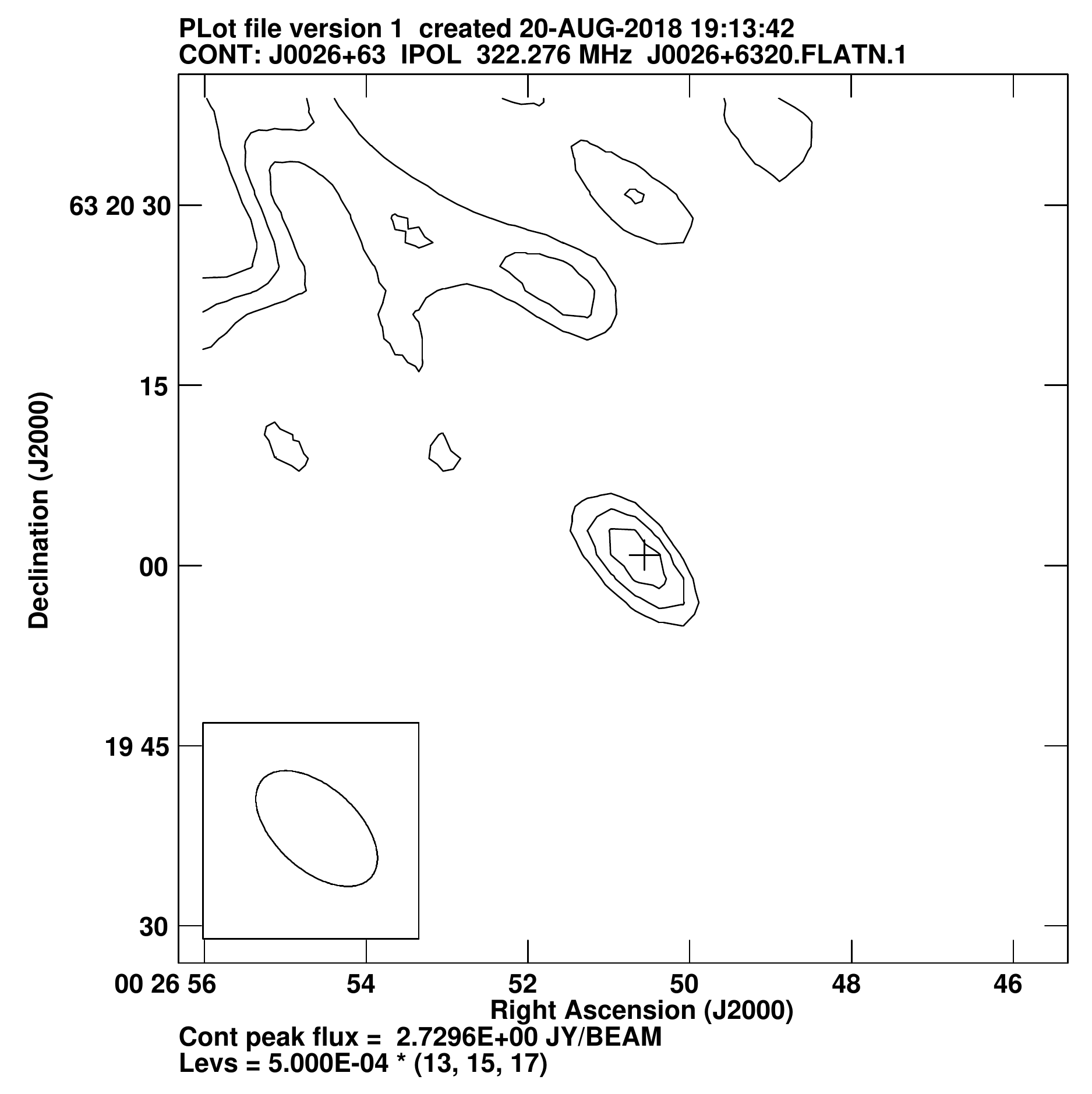}
\includegraphics[trim=0.5cm 1.2cm 0 1.2cm, clip, scale=0.315]{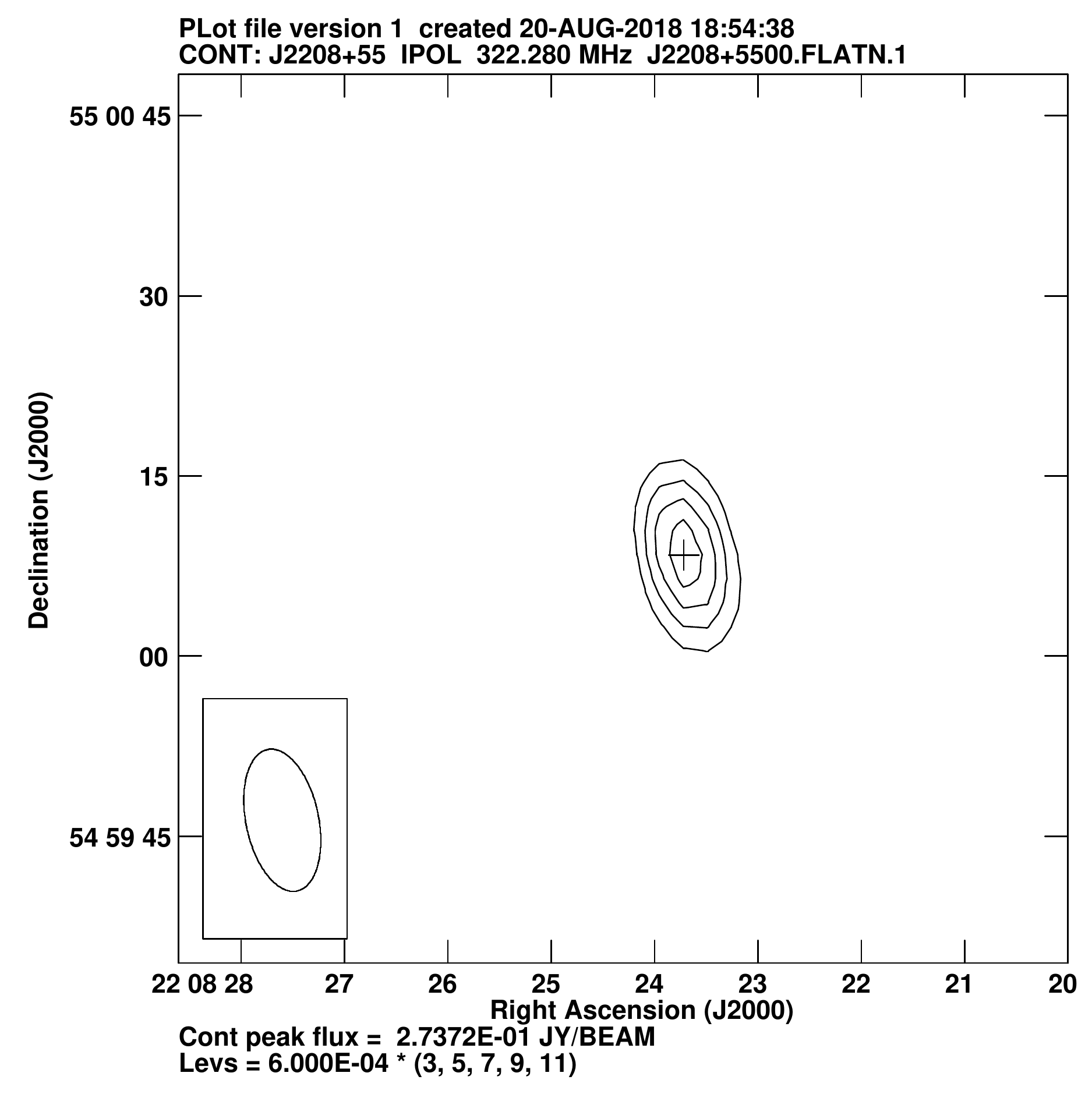}
\includegraphics[trim=0.5cm 1.2cm 0 1.2cm, clip, scale=0.315]{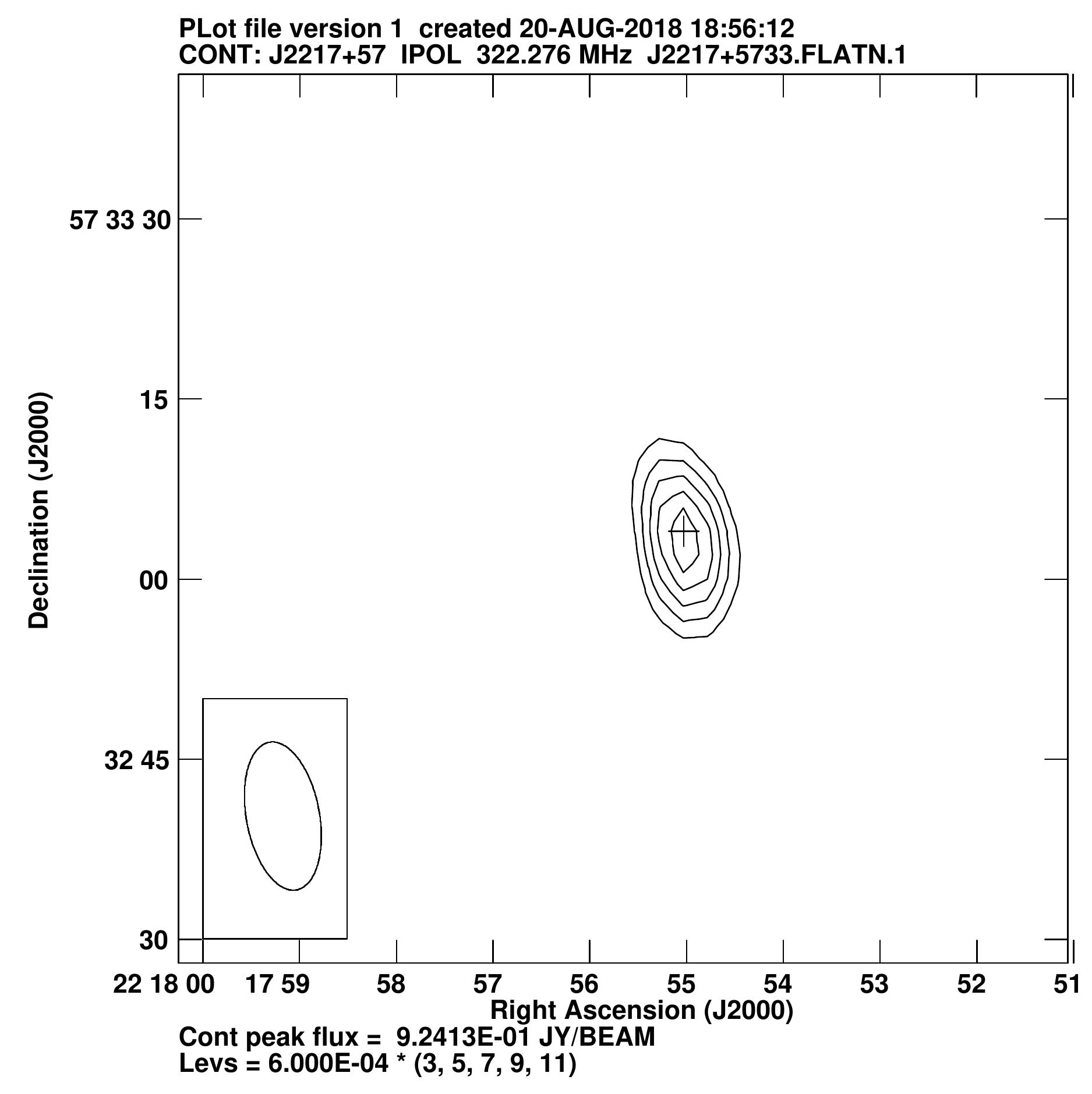}
\caption{Interferometric images of the three pulsars at 325 MHz made with the GMRT. {\it Left:} PSR J0026+6320 with contours at 6.5, 7.5, and 8.5 mJy. The structure seen towards the top left corner in this image is the artifact due to the presence of the Tycho supernova remnant, which is a very strong source. {\it Middle:} PSR J2208+5500 with contours at 1.8, 3.0, 4.2, 5.4, and 6.6 mJy. {\it Right:} PSR J2217+5733 with contours at 1.8, 3.0, 4.2, 5.4, and 6.6 mJy. The synthesized beam is plotted in the bottom left corner in all the panels. The crosses indicate positions of the pulsars from their timing solutions. Note that the error in the timing position (see Table \ref{res_timsol}) is much less than the size of the cross.}
\label{anal_imag_imag}

\end{center}
\end{figure*}

\section{Results}
\label{anal_res}
The results of the data analysis obtained for the three pulsars are presented below in different subsections with each subsection dedicated to one pulsar. Note that the analysis of the imaging data resulted only in the measurements of flux densities and spectral indices, while the rest of the results were obtained from the analysis of the time-series data.

\begin{figure}
\begin{center}

\includegraphics[scale=0.7]{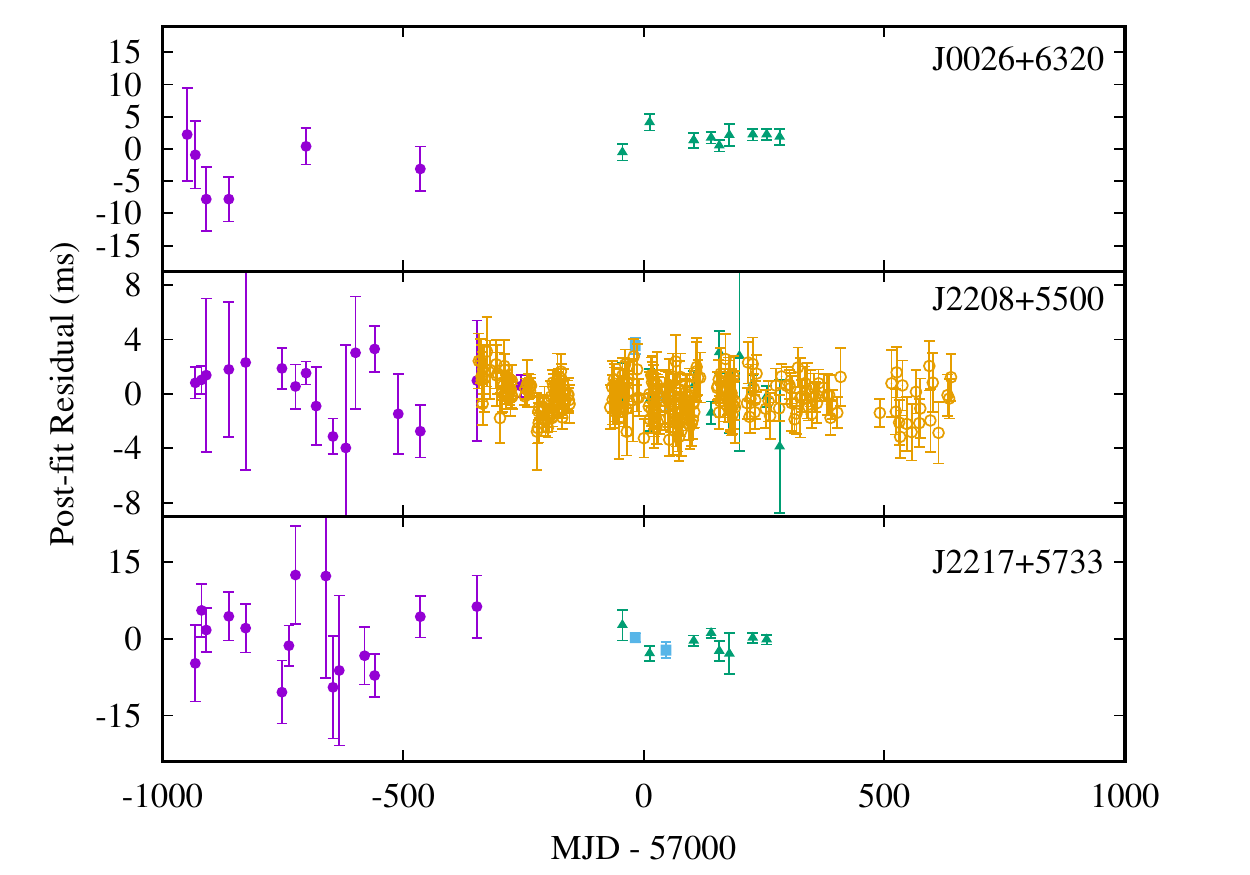}
\caption{Post-fit timing residuals obtained for the pulsars under study. The GMRT TOA measurements are indicated by violet filled circles (325 MHz), dark cyan filled triangles (610 MHz) and light cyan filled squares (1170 MHz), while the ORT TOA measurements are indicated by orange open circles (326 MHz).} 
\label{res_all_timres}

\end{center}
\end{figure}

\begin{table*}

\begin{center}

\begin{tabular}{cccccc}
\hline \hline
Pulsar     & S\subs{1170 MHz} & S\subs{610 MHz} & S\subs{325 MHz} & $\alpha$   \\
           & (mJy)            & (mJy)           & (mJy)           &            \\
\hline
J0026+6320 & -                & 2.9(0.8)        & 8.0(4.0)        & $-$0.9(0.4) \\
J2208+5500 & 0.4(0.1)         & 2.0(0.4)        & 5.8(1.5)        & $-$2.4(0.4) \\
J2217+5733 & 0.5(0.1)         & 4.6(1.1)        & 10(3)           & $-$2.4(0.3) \\
\hline
\end{tabular}

\caption{Flux densities of the observed pulsars at different observing frequencies with 1$\sigma$ errors reported in brackets. We take the rms deviation of the flux densities as the error at 325 MHz, while the errors at other frequencies are measurement errors. Spectral indices reported here were calculated by using these measurements along with the ones obtained by \cite{jml+09}. All our measurements were obtained using imaging observations.}
\label{res_1937_fluxes}
\end{center}

\end{table*}

\begin{figure}
\begin{center}

\includegraphics[scale=0.7]{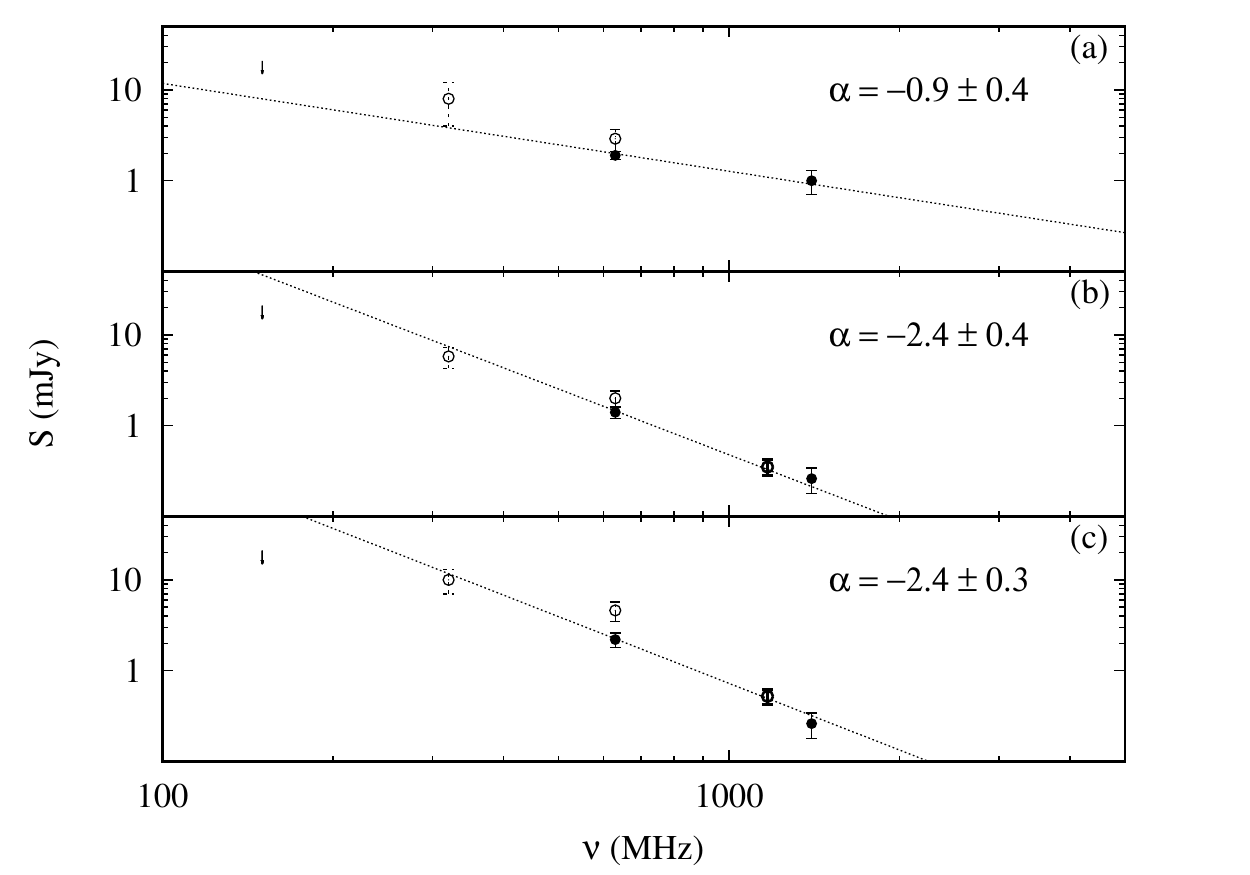}
\caption{Spectra for the pulsars studied in this work. (a): PSR J0026+6320, (b): PSR J2208+5500 and (c): PSR J2217+5733. The filled circles indicate the measurements obtained by \cite{jml+09}, open circles indicate the measurements using images made at 325, 610 and 1170 MHz with GMRT data, while arrows indicate the upper limit obtained from the TGSS \citep{ijm+16}.}
\label{res_spec}

\end{center}
\end{figure}

\subsection{PSR J0026+6320}
\label{res_0026}
The pulse profile for this pulsar shows a single component, which is significantly scatter-broadened at 325 MHz (Figure \ref{res_0026_scatt}). 

The low expected flux density of 1 mJy implied non-detection at 1170 MHz, which was confirmed in observations. Thus, the timing solution was obtained with 15 TOAs from 325 and 610 MHz observations at the GMRT (see Table \ref{res_timsol}). The timing residuals obtained are shown in Figure \ref{res_all_timres}.

The DM of this pulsar is 245 pc cm\sups{--3}. The maximum DM predicted by the NE2001 model \citep{cl02} in this direction is 240 pc cm\sups{--3}, while that predicted by the YMW16 model \citep{ymw17} is 375 pc cm\sups{--3}. Interestingly, one model puts this pulsar right on the edge of the Galaxy, while the other puts it well within. We used our multi-frequency timing analysis to obtain a more refined DM estimate listed in Table \ref{res_timsol}. 

As the S/N for single pulses was poor, it was difficult to identify any null pulses. Additionally, no absence in emission was seen after integrating several pulses. While this implies that no long duration nulls are shown by this pulsar, short period nulls cannot be ruled out. We also did not estimate the modulation index due to poor single-pulse S/N. 

The profile reported by \cite{jml+09} at 1400 MHz was used as a template, or unscattered, profile for estimating $\tau$\subs{sc}, using the method described in \cite{kmn+15}. The profiles for PSR J0026+6320, observed at 325 and 610 MHz, were fitted to obtain estimates of $\tau$\subs{sc}. The fits are shown along with reduced $\chi$\sups{2} in Figure \ref{res_0026_scatt}. The scatter-broadening analysis resulted in $\tau$\subs{sc} of 7.5(2.7) ms at 610 MHz and 47.6(0.6) ms at 325 MHz. These results imply a scatter-broadening frequency scaling index of $-$2.9(1.0), which is flatter than $-$4.4, which is expected for a Kolmogorov spectrum of interstellar medium (ISM) density perturbations. The small values of reduced $\chi$\sups{2} (see Figure \ref{res_0026_scatt}) could be due to overestimation of errors caused by baseline variation in the off-pulse region. Hence, we quote our estimates with three standard deviation errors \citep[see][for more details]{kmn+15} for this pulsar. 

The mean flux densities for PSR J0026+6320 are listed in Table \ref{res_1937_fluxes}. Using these flux densities along with the ones measured by \cite{jml+09}, the spectral index was calculated to be $-$0.9(0.4) (as shown in Figure \ref{res_spec}), which is consistent with the value of $-$0.8(0.3) obtained by \cite{jml+09}. The non-detection as well as the flux density upper limit from TGSS ADR (shown by an arrow in Figure \ref{res_spec}) at 150 MHz is also consistent with expectations.

\subsection{PSR J2208+5500}
\label{res_2208}
The average profiles of PSR J2208+5500, shown in Figure \ref{anal_tim_2208}, consist of a narrow single component which broadens with decreasing frequency. Pulsed emission was detected at all our observing frequencies for this pulsar. The phase-coherent timing solution is listed in Table \ref{res_timsol}. It is consistent with the one obtained by \cite{jml+09}. The post-fit timing residuals are shown in Figure \ref{res_all_timres}. As is clearly seen, the simultaneous TOA measurements obtained at GMRT and ORT agree well with each other, after taking into account the instrumental offset measurements. Within our precision, no significant DM variation is evident after fitting a piecewise-constant model (DMX) with a time window of one week. No significant scatter-broadening was obtained after fitting.

\begin{figure}
\begin{center}

\includegraphics[angle=0,scale=0.35]{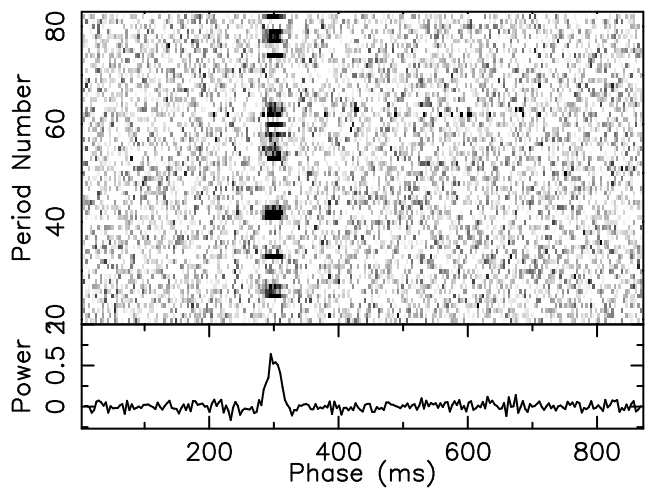}
\caption{Pulse nulling seen in PSR J2208+5500 at 325 MHz. The upper panel shows a gray-scale plot of pulse number vs pulse phase in bins. The bottom panel is the integrated pulse profile (in arbitrary units). Five prominent nulls are clearly seen.}
\label{res_2208_nulling}

\end{center}
\end{figure}

\begin{figure}
\begin{center}

\includegraphics[scale=0.35]{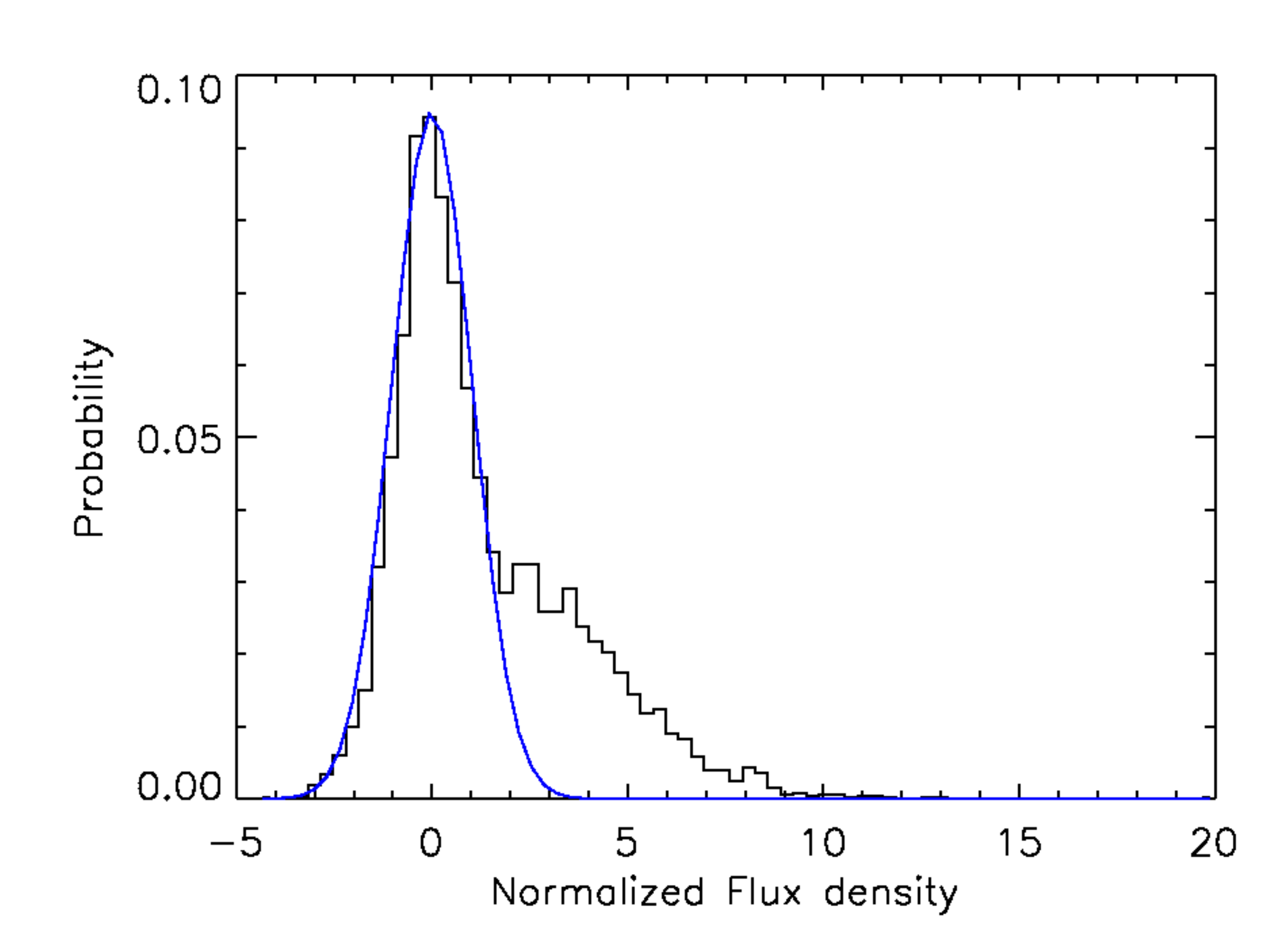}
\caption{Distribution of on-pulse energies for PSR J2208+5500. Solid blue line shows a scaled version of the off-pulse energy distribution fitted to the on-pulse distribution.}
\label{res_2208_hist}

\end{center}
\end{figure}

\cite{jml+09} have suggested that PSR J2208+5500 might be a nulling pulsar from their observations carried out with the GMRT at 610 MHz. Due to the steep spectral index of this pulsar and the use of the incoherent array, they did not have sufficient S/N for single pulses and hence, could estimate only a lower limit of 7.5\% on the NF. On the other hand, we have observed at 325 MHz with the sensitive PA mode of the GMRT. This has provided an average single-pulse S/N of three and has enabled us to obtain a better estimate of the NF (Figure \ref{res_2208_nulling} and \ref{res_2208_hist}).

The nulling analysis was performed using the method devised by \cite{r76}. Data taken at 10 separate epochs were selected for the nulling analysis as they had single-pulse S/N $>$ 3 due to smaller amounts of RFI. The analysis was performed on data taken at each epoch independently. The de-dispersed time-series was folded modulo the pulse period to 256 bins and dumped every period into a single-pulse file. Periods affected by RFI were visually identified and masked. Two windows of equal width were visually identified for each pulse. The window with phase bins where the pulsed emission is present is called the on-pulse window (e.g. bin number 75$-$90 in Figure \ref{res_2208_nulling}), whereas a window away from the pulsed emission is called the off-pulse window (e.g. bin number 150$-$165 in Figure \ref{res_2208_nulling}). The on-pulse window was selected such that it equaled the pulse width at 10\% of the peak intensity in the integrated profile. The integrated energies in the on-pulse and the off-pulse windows for each rotation period were normalized with the mean pulse energy \citep[see][for more details on the analysis technique]{gjk12} and combined to form a single distribution of normalized on-pulse and off-pulse energies. The combined data provided an ensemble of 5251 pulse periods. The pulse-energy distributions are shown in Figure \ref{res_2208_hist}. The on-pulse energy histogram shows a clear excess at zero mean, which corresponds to the null pulses. A fit by the modeled off-pulse energy distribution to the on-pulse energy distribution is also shown. The NF, given by the fraction of the off-pulse distribution required to fit the zero mean excess in the on-pulse distribution, is estimated to be 53(3)\%. The apparent NF usually depends on many factors, such as the true NF, observing time per epoch, number of epochs and their separation etc. We had individual observations of 10 minutes duration but they were spread over a time span of almost three years. In order to see how the apparent NF changes with number of accumulated pulses, we also estimated the NF for each of the individual observing epochs (10 in all with typically 550 pulses each). The NF varied from 35 to 60\% for individual epochs. Thus, to get a more accurate NF, we formed 10 random sequences of the NFs and estimated the cumulative NF for each sequence to see how the changing apparent NF at individual epochs affects the estimation of the true NF from the full data set (i.e. all the pulses from 10 epochs stacked together). This analysis showed that although the apparent NF varies from epoch to epoch, the cumulative NF actually converges to 53\% within the 3\% error after adding roughly 6--7 epochs together. Thus, we believe that our estimate of the NF is quite robust.

\begin{table*}
\begin{center}
\begin{scriptsize}
\caption{Timing solutions obtained for all three pulsars. For the measured quantities, numbers in brackets indicate weighted 1$\sigma$ errors in the last significant digit as reported by TEMPO2. For the set quantities, numbers in brackets indicate 1$\sigma$ errors in the last significant digit [as reported by \cite{jml+09}].}
\label{res_timsol}
\begin{tabular}{lccc}
\hline\hline
\multicolumn{4}{c}{Fit and dataset} \\
\hline
Pulsar name\dotfill & J0026+6320 & J2208+5500 & J2217+5733 \\
MJD range\dotfill & 56051---57283 & 56051---57283 & 56051---57283 \\
Data span (yr)\dotfill & 3.4 & 3.4 & 3.4 \\
Number of TOAs\dotfill & 15 & 164 & 29 \\
RMS timing residual (ms)\dotfill & 1.6 & 1.2 & 1.9 \\
\hline
\multicolumn{4}{c}{Measured Quantities} \\
\hline
Spin frequency, $\nu$ (s$^{-1}$)\dotfill & 3.1411201226(2) & 1.07162470101(3) & 0.94621311300(3) \\
First derivative of spin frequency, $\dot{\nu}$ (s$^{-2}$)\dotfill & $-$1.486(5)$\times 10^{-15}$ & $-$8.023(3)$\times 10^{-15}$ & $-$5.79(1)$\times 10^{-16}$ \\
Dispersion measure, DM (cm$^{-3}$pc)\dotfill & 245.06(6) & 105.21(1) & 130.67(6) \\
Epoch (MJD)\dotfill & 56675 & 56675 & 56675 \\
\hline
\multicolumn{4}{c}{Set Quantities} \\
\hline
Right ascension, $\alpha$ (hh:mm:ss)\dotfill & 00:26:50.561(8) & 22:08:23.72(1) & 22:17:55.03(2) \\
Declination, $\delta$ (dd:mm:ss)\dotfill & +63:20:00.87(5) & +55:00:08.42(5) & +57:33:04.0(2) \\
Galactic longitude, $l$ (deg)\dotfill & 120.17644(3) & 100.93858(4) & 103.47196(8) \\
Galactic latitude, $b$ (deg)\dotfill & +0.59367(1) & $-$0.75066(1) & +0.59943(6) \\
\hline
\multicolumn{4}{c}{Derived Quantities} \\
\hline
$\log_{10}$(Characteristic age, yr) \dotfill & 7.5 & 6.3 & 7.4 \\
$\log_{10}$(Surface magnetic field strength, G) \dotfill & 11.4 & 12.4 & 11.9 \\
$\log_{10}$(\.{E}, ergs/s) \dotfill & 32.3 & 32.5 & 31.3 \\
\hline
\multicolumn{4}{c}{Assumptions} \\
\hline
Clock correction procedure\dotfill & \multicolumn{3}{c}{TT(TAI)} \\
Solar system ephemeris model\dotfill & \multicolumn{3}{c}{DE405} \\
Binary model\dotfill & \multicolumn{3}{c}{NONE} \\
Model version number\dotfill & \multicolumn{3}{c}{5.00} \\
\hline
\end{tabular}
\end{scriptsize}
\end{center}
\end{table*}

The mean flux densities of PSR J2208+5500 are listed in Table \ref{res_1937_fluxes}. Using these flux densities and the ones measured by \cite{jml+09}, the spectral index was calculated to be $-$2.4(0.4) (as shown in Figure \ref{res_spec}), which is consistent with the value of $-$2.0(0.9) obtained by \cite{jml+09}. The upper limit from TGSS ADR (shown by an arrow in Figure \ref{res_spec}) most likely indicates a spectral break or a turnover below 300 MHz.

\subsection{PSR J2217+5733}
\label{res_2217}

The phase-coherent timing solution for this pulsar using the GMRT TOAs is listed in Table \ref{res_timsol} and the post-fit residuals are shown in Figure \ref{res_all_timres}.

A refined DM estimate was obtained using the multi-frequency timing analysis and is reported in Table \ref{res_timsol}. As the pulsar is weaker than PSR J2208+5500, a single-pulse study was not possible. We see no evidence of pulse nulling after integrating up to 30 pulses, which does not preclude shorter nulls. 

PSR J2217+5733 was observed at 325, 610, and 1170 MHz. Significant scatter-broadening is seen at 325 MHz (see Figure \ref{res_2217_scatt}). The profile reported by \cite{jml+09} at 1400 MHz was used as a template profile to estimate $\tau$\subs{sc}, using the method described in \cite{kmn+15}. Our analysis resulted in $\tau$\subs{sc} of 9.7(1.2) ms at 610 MHz and 61.9(3.3) ms at 325 MHz. It is worth noting that these values are more than five times the values predicted by the NE2001 model \citep{cl02} and there seems to be no apparent diffuse emission associated with this pulsar (see right panel of Figure \ref{anal_imag_imag}).
These fits and the reduced $\chi$\sups{2} are shown in Figure \ref{res_2217_scatt} and imply a scatter-broadening frequency scaling index of $-$2.9(0.5). This is much flatter than  $-$4.4, expected for a Kolmogorov spectrum of ISM density perturbations.

The mean flux densities of PSR J2217+5733 are listed in Table \ref{res_1937_fluxes}. Using these flux densities along with the ones measured by \cite{jml+09}, the spectral index was calculated to be $-$2.4(0.3) (as shown in Figure \ref{res_spec}), which is consistent with the value of $-$2.7(1.3) obtained by \cite{jml+09}. The upper limit from TGSS ADR (shown by an arrow in Figure \ref{res_spec}) most likely indicates a spectral break or a turnover below 300 MHz for this pulsar as well.

\section{Discussion}
\label{disc}

\cite{jml+09} suspected that PSR J2208+5500 shows nulls but did not have enough single-pulse S/N in their incoherent array observations at 610 MHz to estimate its NF. Our phased array observations at 325 MHz suggest very significant nulling (NF $\sim$53\%) with the pattern of nulling very similar to pulsars, such as PSR B1112+50 \citep[NF $\sim$60\%;][]{gjk12} consisting of bursts and nulls with a similar duration (1 to 10 pulses). At the same time, this pattern differs from that shown by PSR B2034+19, a pulsar with a nulling fraction of 44\% \citep{hr09}, but showing different null and burst lengths with nulls of up to four pulses and bursts as large as 20$-$30 pulses. This suggests that pulsars with similar NF show different nulling patterns, as reported earlier by \cite{gjk12}. We also note from our multi-epoch NFs as well as cumulative NFs that the NF estimated from a single-epoch observation is highly dependent on observing conditions and nulling pattern of the pulsar.

The average profile of PSR J2208+5500 has a single component (Figure \ref{anal_tim_2208}) from 325 to 1175 MHz, showing little evolution. At the same time, the spectral index of the emission is steep ($-$2.4). This suggests that this component is a core component, as the spectral indices of core components are expected to be much steeper than the conal outriders \citep{p02}. This is seen in many pulsars (e.g. PSRs B0329+54, B1642$-$03 and B0355+54) and was confirmed in a systematic study of pulsar beams \cite[][see Figure 10]{lm88}. The half-power width of the profile for PSR J2208+5500 is about 8$^\circ$, which is comparable to the typical width of core components \citep{r90}. It is also much smaller than the typical width of conal single profiles \citep{lm88}. The width-period relation for core-single pulsars \citep{r90} gives an expected core width of 5.4$^\circ$ for an orthogonal rotator, close to that observed. If an orthogonal rotator is not assumed, the same relation suggests an inclination angle of 42$^\circ$ for this pulsar. Interestingly, the ratio of the surface magnetic field and the period of the pulsar (B$_{12}$/P) has a value of 2.75, typical of the core single class of pulsars \citep{r90}. This is much larger than that expected for the conal single class of pulsars. 

While we did not have polarimetric observations to confirm these implications, the above arguments suggest that the single component seen in the pulsar is a core component. If indeed this component is a core component, then this is unusual because core single pulsars typically do not show nulling \citep{r86,r90}. To the best of our knowledge, it is the only pulsar known so far showing significant NF in a core component \citep[see][for a recent study]{bmm17}. This is inconsistent with a geometric origin for nulls due to `empty' sightline traverses, as was proposed for few pulsars \citep{dr01}, and implies that nulling is more likely due to intrinsic changes in the pulsar magnetosphere. Assuming that the core emission switches off during the nulls, this pulsar could provide a link between nulling pulsars and intermittent pulsars where the pulsed emission is quenched for much longer timescales \citep[e.g.][]{llm+12}. Future full polarization observations as well as high-sensitivity low-frequency observations to detect sub-pulse drift are motivated by our results. If drifting is detected in these observations, we can rule out the possibility that the single component in this pulsar is a core component as sub-pulse drifting always denotes a conal component \citep{r86}.

Simultaneous imaging data acquired at the GMRT during the timing observations were helpful in determining the spectral indices through well-calibrated flux densities. These measurements together with the 150 MHz flux densities from the TGSS ADR suggest a spectral break or turnover between 100$-$300 MHz for PSRs J2208+5500 and J2217+5733 and a possible spectral break for PSR J0026+6320 below 300 MHz. Spectral breaks and turnovers are important for the understanding of emission mechanism of radio pulsars. The turnovers could be at high frequencies (of the order of 1 GHz) due to absorption in the adjacent or intervening medium \citep{lrk+15,rla16} or at low radio frequencies due to the distribution of the particle energies responsible for the radiation or it could be due to synchrotron self-absorption in the magnetosphere. The origin of the low frequency turnover is not entirely clear due to lack of flux density measurements at low radio frequencies \citep{fjm+16}. Future observations of these pulsars with the GMRT at 150 MHz with deeper integration than the TGSS are motivated in order to either detect these pulsars or put better upper limits on their flux densities at 150 MHz. This will better constrain the turnover frequency. It may be noted that none of these pulsars are associated with any wind nebula or supernova remnant (see Figure \ref{anal_imag_imag}). 

Hence, the low frequency turnover in these pulsars could be magnetospheric in nature. This could provide an independent handle on the density of the magnetospheric plasma which is responsible for the coherent radio emission. Future observations between 100--300 MHz will be useful to confirm this picture.

The frequency scaling indices for PSRs J0026 +6320 and J2217+5733 are not consistent with the expected Kolmogorov spectrum for turbulence, as also reported in some recent studies \citep{lmg+04,kmn+15,kjm17}. Multi-path propagation from multiple screens is proposed as a plausible explanation for this deviation in these studies. Our results further enhance the sample of pulsars showing such deviations for future detailed modeling of the ISM. With a scatter-broadening frequency scaling index of $-$2.9, the YMW16 model \citep{ymw17} yields an expected $\tau_{sc}$ for PSR J0026+6320 of about 4 and 26 ms at 610 and 325 MHz, respectively. The estimates for PSR J2217+5733 are 0.4 and 2.6 ms at 610 and 325 MHz, respectively. The large scatter-broadening seen in these two pulsars compared to the predictions [although these measurements lie within the scatter of the $\tau_{sc}$ measurements at low frequencies \citep[e.g.][]{kmn+15}] and the fact that these pulsars were discovered at 610 MHz, suggest that both high-frequency and low-frequency pulsar searches discriminate against weak and distant pulsars. Thus, there is a selection effect in these surveys, which needs to be taken into account in pulsar population models. Furthermore, these results suggest that a Galactic plane pulsar survey at an intermediate frequency range, such as 550$-$950 MHz, has a potential to discover many distant pulsars missed by both high-frequency and low-frequency pulsar surveys. The Five hundred metre Aperture Spherical Telescope in China has already shown good success with a pulsar survey\footnote{See http://crafts.bao.ac.cn/pulsar/ for more details.} in a similar frequency range. Such a survey will be possible in future with instruments such as the upgraded GMRT and Square Kilometre Array and our results strongly motivate such surveys.  

Pulsar searches typically lead to a coarse estimate of the pulsar DM, particularly at higher frequencies (such as 610 MHz in the case of these pulsars). With our multi-frequency observations employing frequencies as low as 325 MHz, the inverse square dependence of the dispersion smearing with frequency allowed determination of a much better measurement of the pulsar DM with pulsar timing techniques for all three pulsars. This underlines the importance of low frequency (down to 325 MHz) follow-up observations for high precision DM estimates of millisecond pulsars discovered at high radio frequencies. 

Simultaneous and quasi-simultaneous multi-frequency observations carried out with the GMRT and the ORT have determined the coherent timing solution for PSR J2208+5500 and demonstrated that even with a single polarization detector, the ORT is capable of providing high quality timing data with DM determination of the order of 0.01 pc cm\sups{--3}. This was achieved only with a very few TOAs obtained at frequencies higher than 325 MHz with typical TOA errors of the order of a few hundred $\mu$s. In fact, the DM correction for PSR J2208+5500 (a change of 5\% as compared to the search DM) could be obtained only after multi-frequency timing analysis and not otherwise. The natural extension of this experiment would be to have a timing program for some of the millisecond pulsars which are a part of the international pulsar timing array\footnote{http://www.ipta4gw.org}. For these pulsars, with single-epoch simultaneous ORT (325 MHz) and GMRT (1300 MHz) observations with the typical TOA errors of the order of a few $\mu$s, the DMs could be determined with an accuracy of up to 0.0001 pc cm\sups{--3} which is comparable to the precision currently achieved by other pulsar timing array experiments \citep[e.g.][]{abb+18}. 

The success of such an experiment would prove to be useful for the timing array experiments as the ORT can provide TOAs at a much higher cadence (weekly or bi-weekly) and provide complementary data at 325 MHz, which track the DM variations over time scales shorter than the cadence of 2--4 weeks typically used for pulsar timing arrays \citep[e.g.][]{mhb+13,dcl+16,jml+17}.

\section{Acknowledgments}
\label{ack}
We thank the staff of the ORT and the GMRT, who have made these observations possible. The ORT and the GMRT are operated by the National Centre for Radio Astrophysics of the Tata Institute of Fundamental Research. MPS would like to thank Ishwara Chandra, and Ruta Kale for the help regarding advanced imaging techniques. MAM and MPS are supported by NSF award number OIA--1458952. MAM and MPS are members of the NANOGrav Physics Frontiers Center which is supported by NSF award 1430284. This work made use of the PONDER receiver, which was funded by XII plan grant TIFR 12P0714. BCJ, PKM and MAK also acknowledge support for this work from DST-SERB grant EMR/2015/000515.

\bibliography{ref}

\begin{thebibliography}{}
\expandafter\ifx\csname natexlab\endcsname\relax\def\natexlab#1{#1}\fi

\bibitem[{{Arzoumanian} {et~al.}(2018){Arzoumanian}, {Brazier},
  {Burke-Spolaor}, {Chamberlin}, {Chatterjee}, {Christy}, {Cordes}, {Cornish},
  {Crawford}, {Thankful Cromartie}, {Crowter}, {DeCesar}, {Demorest}, {Dolch},
  {Ellis}, {Ferdman}, {Ferrara}, {Fonseca}, {Garver-Daniels}, {Gentile},
  {Halmrast}, {Huerta}, {Jenet}, {Jessup}, {Jones}, {Jones}, {Kaplan}, {Lam},
  {Lazio}, {Levin}, {Lommen}, {Lorimer}, {Luo}, {Lynch}, {Madison}, {Matthews},
  {McLaughlin}, {McWilliams}, {Mingarelli}, {Ng}, {Nice}, {Pennucci}, {Ransom},
  {Ray}, {Siemens}, {Simon}, {Spiewak}, {Stairs}, {Stinebring}, {Stovall},
  {Swiggum}, {Taylor}, {Vallisneri}, {van Haasteren}, {Vigeland}, {Zhu}, \&
  {The NANOGrav Collaboration}}]{abb+18}
{Arzoumanian}, Z., {Brazier}, A., {Burke-Spolaor}, S., {et~al.} 2018, \apjs,
  235, 37

\bibitem[{{Basu} {et~al.}(2017){Basu}, {Mitra}, \& {Melikidze}}]{bmm17}
{Basu}, R., {Mitra}, D., \& {Melikidze}, G.~I. 2017, \apj, 846, 109

\bibitem[{{Bates} {et~al.}(2013){Bates}, {Lorimer}, \& {Verbiest}}]{blv13}
{Bates}, S.~D., {Lorimer}, D.~R., \& {Verbiest}, J.~P.~W. 2013, \mnras, 431,
  1352

\bibitem[{{Bilous} {et~al.}(2016){Bilous}, {Kondratiev}, {Kramer}, {Keane},
  {Hessels}, {Stappers}, {Malofeev}, {Sobey}, {Breton}, {Cooper}, {Falcke},
  {Karastergiou}, {Michilli}, {Os{\l}owski}, {Sanidas}, {ter Veen}, {van
  Leeuwen}, {Verbiest}, {Weltevrede}, {Zarka}, {Grie{\ss}meier}, {Serylak},
  {Bell}, {Broderick}, {Eisl{\"o}ffel}, {Markoff}, \& {Rowlinson}}]{bkk+16}
{Bilous}, A.~V., {Kondratiev}, V.~I., {Kramer}, M., {et~al.} 2016, \aap, 591,
  A134

\bibitem[{Chengalur(2013)}]{c13}
Chengalur, J.~N. 2013, {FLAGCAL}: A flagging and calibration pipeline for
  {GMRT} data, Tech. Rep. R258, National Centre for Radio Astrophysics,
  Ganeshkhind, Pune

\bibitem[{{Condon} {et~al.}(1998){Condon}, {Cotton}, {Greisen}, {Yin},
  {Perley}, {Taylor}, \& {Broderick}}]{ccg+98}
{Condon}, J.~J., {Cotton}, W.~D., {Greisen}, E.~W., {et~al.} 1998, \aj, 115,
  1693

\bibitem[{{Cordes} \& {Lazio}(2002)}]{cl02}
{Cordes}, J.~M., \& {Lazio}, T.~J.~W. 2002, ArXiv Astrophysics e-prints,
  astro-ph/0207156

\bibitem[{{Dembska} {et~al.}(2015){Dembska}, {Basu}, {Kijak}, \&
  {Lewandowski}}]{dbk+15}
{Dembska}, M., {Basu}, R., {Kijak}, J., \& {Lewandowski}, W. 2015, \mnras, 449,
  1869

\bibitem[{{Deshpande} \& {Rankin}(2001)}]{dr01}
{Deshpande}, A.~A., \& {Rankin}, J.~M. 2001, \mnras, 322, 438

\bibitem[{{Desvignes} {et~al.}(2016){Desvignes}, {Caballero}, {Lentati},
  {Verbiest}, {Champion}, {Stappers}, {Janssen}, {Lazarus}, {Os{\l}owski},
  {Babak}, {Bassa}, {Brem}, {Burgay}, {Cognard}, {Gair}, {Graikou},
  {Guillemot}, {Hessels}, {Jessner}, {Jordan}, {Karuppusamy}, {Kramer},
  {Lassus}, {Lazaridis}, {Lee}, {Liu}, {Lyne}, {McKee}, {Mingarelli},
  {Perrodin}, {Petiteau}, {Possenti}, {Purver}, {Rosado}, {Sanidas}, {Sesana},
  {Shaifullah}, {Smits}, {Taylor}, {Theureau}, {Tiburzi}, {van Haasteren}, \&
  {Vecchio}}]{dcl+16}
{Desvignes}, G., {Caballero}, R.~N., {Lentati}, L., {et~al.} 2016, \mnras, 458,
  3341

\bibitem[{{Frail} {et~al.}(2016){Frail}, {Jagannathan}, {Mooley}, \&
  {Intema}}]{fjm+16}
{Frail}, D.~A., {Jagannathan}, P., {Mooley}, K.~P., \& {Intema}, H.~T. 2016,
  \apj, 829, 119

\bibitem[{{Gajjar} {et~al.}(2012){Gajjar}, {Joshi}, \& {Kramer}}]{gjk12}
{Gajjar}, V., {Joshi}, B.~C., \& {Kramer}, M. 2012, \mnras, 424, 1197

\bibitem[{{Herfindal} \& {Rankin}(2009)}]{hr09}
{Herfindal}, J.~L., \& {Rankin}, J.~M. 2009, \mnras, 393, 1391

\bibitem[{{Hobbs} {et~al.}(2006){Hobbs}, {Edwards}, \& {Manchester}}]{hem06}
{Hobbs}, G.~B., {Edwards}, R.~T., \& {Manchester}, R.~N. 2006, \mnras, 369, 655

\bibitem[{{Intema} {et~al.}(2017){Intema}, {Jagannathan}, {Mooley}, \&
  {Frail}}]{ijm+16}
{Intema}, H.~T., {Jagannathan}, P., {Mooley}, K.~P., \& {Frail}, D.~A. 2017,
  \aap, 598, A78

\bibitem[{{Jones} {et~al.}(2017){Jones}, {McLaughlin}, {Lam}, {Cordes},
  {Levin}, {Chatterjee}, {Arzoumanian}, {Crowter}, {Demorest}, {Dolch},
  {Ellis}, {Ferdman}, {Fonseca}, {Gonzalez}, {Jones}, {Lazio}, {Nice},
  {Pennucci}, {Ransom}, {Stinebring}, {Stairs}, {Stovall}, {Swiggum}, \&
  {Zhu}}]{jml+17}
{Jones}, M.~L., {McLaughlin}, M.~A., {Lam}, M.~T., {et~al.} 2017, \apj, 841,
  125

\bibitem[{{Joshi} {et~al.}(2009){Joshi}, {McLaughlin}, {Lyne}, {Ludovici},
  {Pawar}, {Faulkner}, {Lorimer}, {Kramer}, \& {Davies}}]{jml+09}
{Joshi}, B.~C., {McLaughlin}, M.~A., {Lyne}, A.~G., {et~al.} 2009, \mnras, 398,
  943

\bibitem[{{Keith} {et~al.}(2013){Keith}, {Coles}, {Shannon}, {Hobbs},
  {Manchester}, {Bailes}, {Bhat}, {Burke-Spolaor}, {Champion}, {Chaudhary},
  {Hotan}, {Khoo}, {Kocz}, {Os{\l}owski}, {Ravi}, {Reynolds}, {Sarkissian},
  {van Straten}, \& {Yardley}}]{kcs+13}
{Keith}, M.~J., {Coles}, W., {Shannon}, R.~M., {et~al.} 2013, \mnras, 429, 2161

\bibitem[{{Krishnakumar} {et~al.}(2017){Krishnakumar}, {Joshi}, \&
  {Manoharan}}]{kjm17}
{Krishnakumar}, M.~A., {Joshi}, B.~C., \& {Manoharan}, P.~K. 2017, \apj, 846,
  104

\bibitem[{{Krishnakumar} {et~al.}(2015){Krishnakumar}, {Mitra}, {Naidu},
  {Joshi}, \& {Manoharan}}]{kmn+15}
{Krishnakumar}, M.~A., {Mitra}, D., {Naidu}, A., {Joshi}, B.~C., \&
  {Manoharan}, P.~K. 2015, \apj, 804, 23

\bibitem[{{Lam} {et~al.}(2016){Lam}, {Cordes}, {Chatterjee}, {Jones},
  {McLaughlin}, \& {Armstrong}}]{lcc+16}
{Lam}, M.~T., {Cordes}, J.~M., {Chatterjee}, S., {et~al.} 2016, \apj, 821, 66

\bibitem[{{Lewandowski} {et~al.}(2015){Lewandowski}, {Ro{\.z}ko}, {Kijak}, \&
  {Melikidze}}]{lrk+15}
{Lewandowski}, W., {Ro{\.z}ko}, K., {Kijak}, J., \& {Melikidze}, G.~I. 2015,
  \apj, 808, 18

\bibitem[{{L{\"o}hmer} {et~al.}(2004){L{\"o}hmer}, {Mitra}, {Gupta}, {Kramer},
  \& {Ahuja}}]{lmg+04}
{L{\"o}hmer}, O., {Mitra}, D., {Gupta}, Y., {Kramer}, M., \& {Ahuja}, A. 2004,
  \aap, 425, 569

\bibitem[{{Lorimer}(2011)}]{sigproc}
{Lorimer}, D.~R. 2011, {SIGPROC: Pulsar Signal Processing Programs},
  Astrophysics Source Code Library, , , ascl:1107.016

\bibitem[{{Lorimer} {et~al.}(2012){Lorimer}, {Lyne}, {McLaughlin}, {Kramer},
  {Pavlov}, \& {Chang}}]{llm+12}
{Lorimer}, D.~R., {Lyne}, A.~G., {McLaughlin}, M.~A., {et~al.} 2012, \apj, 758,
  141

\bibitem[{{Lyne} \& {Manchester}(1988)}]{lm88}
{Lyne}, A.~G., \& {Manchester}, R.~N. 1988, \mnras, 234, 477

\bibitem[{{Manchester} {et~al.}(2013){Manchester}, {Hobbs}, {Bailes}, {Coles},
  {van Straten}, {Keith}, {Shannon}, {Bhat}, {Brown}, {Burke-Spolaor},
  {Champion}, {Chaudhary}, {Edwards}, {Hampson}, {Hotan}, {Jameson}, {Jenet},
  {Kesteven}, {Khoo}, {Kocz}, {Maciesiak}, {Oslowski}, {Ravi}, {Reynolds},
  {Sarkissian}, {Verbiest}, {Wen}, {Wilson}, {Yardley}, {Yan}, \&
  {You}}]{mhb+13}
{Manchester}, R.~N., {Hobbs}, G., {Bailes}, M., {et~al.} 2013, \pasa, 30, e017

\bibitem[{{Naidu} {et~al.}(2015){Naidu}, {Joshi}, {Manoharan}, \&
  {Krishnakumar}}]{njm+15}
{Naidu}, A., {Joshi}, B.~C., {Manoharan}, P.~K., \& {Krishnakumar}, M.~A. 2015,
  Experimental Astronomy, 39, 319

\bibitem[{{Petrova}(2002)}]{p02}
{Petrova}, S.~A. 2002, \aap, 383, 1067

\bibitem[{{Rajwade} {et~al.}(2016){Rajwade}, {Lorimer}, \& {Anderson}}]{rla16}
{Rajwade}, K., {Lorimer}, D.~R., \& {Anderson}, L.~D. 2016, \mnras, 455, 493

\bibitem[{{Rankin}(1986)}]{r86}
{Rankin}, J.~M. 1986, \apj, 301, 901

\bibitem[{{Rankin}(1990)}]{r90}
---. 1990, \apj, 352, 247

\bibitem[{{Ritchings}(1976)}]{r76}
{Ritchings}, R.~T. 1976, \mnras, 176, 249

\bibitem[{{Roy} {et~al.}(2010){Roy}, {Gupta}, {Pen}, {Peterson}, {Kudale}, \&
  {Kodilkar}}]{rgp+10}
{Roy}, J., {Gupta}, Y., {Pen}, U.-L., {et~al.} 2010, Experimental Astronomy,
  28, 25

\bibitem[{{Swarup} {et~al.}(1991){Swarup}, {Ananthakrishnan}, {Kapahi}, {Rao},
  {Subrahmanya}, \& {Kulkarni}}]{sak+91}
{Swarup}, G., {Ananthakrishnan}, S., {Kapahi}, V.~K., {et~al.} 1991, Current
  Science, Vol.~60, NO.2/JAN25, P.~95, 1991, 60, 95

\bibitem[{{Taylor}(1992)}]{t92}
{Taylor}, J.~H. 1992, Royal Society of London Philosophical Transactions Series
  A, 341, 117

\bibitem[{{Yao} {et~al.}(2017){Yao}, {Manchester}, \& {Wang}}]{ymw17}
{Yao}, J.~M., {Manchester}, R.~N., \& {Wang}, N. 2017, \apj, 835, 29

\bibitem[{{You} {et~al.}(2007{\natexlab{a}}){You}, {Hobbs}, {Coles},
  {Manchester}, \& {Han}}]{yhc+07a}
{You}, X.~P., {Hobbs}, G.~B., {Coles}, W.~A., {Manchester}, R.~N., \& {Han},
  J.~L. 2007{\natexlab{a}}, \apj, 671, 907

\bibitem[{{You} {et~al.}(2007{\natexlab{b}}){You}, {Hobbs}, {Coles},
  {Manchester}, {Edwards}, {Bailes}, {Sarkissian}, {Verbiest}, {van Straten},
  {Hotan}, {Ord}, {Jenet}, {Bhat}, \& {Teoh}}]{yhc+07b}
{You}, X.~P., {Hobbs}, G., {Coles}, W.~A., {et~al.} 2007{\natexlab{b}}, \mnras,
  378, 493

\end{thebibliography}
\bibliographystyle{aasjournal}

\end{document}